# Strong equilibration of Landau level's edge-states at the graphene edge

Son T. Le [1,2], Joseph A. Hagmann[2], Nikolai Klimov[2], David Newell[2], Ji Ung Lee[3], Jun Yan[4], Curt A. Richter [2]

[1] *Theiss Research, CA, USA;* [2] *National Institute of Standards and Technology, MD, USA;* [3] *SUNY, NY, USA;* [4] *University of Massachusetts, Amherst, MA, USA*

**Abstract:**

We present the results of an experimental study of the interaction of quantized Landau level (LL) edge-states at the physical edge of graphene by using a graphene *pn* junction device with a ring-shaped geometry for the channel. The unique device geometry allows the interactions between edge-states to be probed at both electrostatic edges defined by *pn* junctions and at the graphene physical edge. Measurements show that while the lowest LL edge-state is decoupled from the other LL's along the electrostatic junction, all the edge-states strongly equilibrate at the graphene physical edge despite the relatively short distance that they travel along the edge in our device. These findings are fundamental for the engineering of future high-performance graphene field-effect transistors based upon electron optics.

**Main text:**

The unique properties of graphene, a single-layer honeycomb lattice of carbon, offer an attractive material platform to study the fundamental physics of high-mobility Dirac fermion electronics and presents the potential for ultra-fast, low-power, high-performance switching devices [1,2,3]. The dual Dirac band structure of graphene gives rise to fundamental properties such as carriers with zero effective mass, suppressed direct backscattering [4], and very high carrier mobility [5,6] that make it appealing for electronic applications. The technical drawback to using graphene as the channel material in a conventional field effect transistors (FETs) is that it behaves as a gapless semimetal which precludes high ON/OFF switching ratios. A proposed alternative switching method in graphene makes use of gate-

Corresponding author: Son T. Le; Email: son.le@nist.gov

tunable transmission gaps across graphene *pn* junctions (G*pn*Js) utilizing the Klein tunneling mechanism [2,3,7]. G*pn*Js angularly sort electrons by longitudinal energy and normal incidence, filtering out off-normal incidence carriers [8,9]. Graphene Klein tunneling FETs (GKTFETs) are based on a sequence of two or more angled *pn*Js. Due to the angular filtering at the *pn*Js interfaces, two junctions at different angles to each other open up a transport-gap which effectively turns the device OFF in the $n^+$-$p$-$n^+$ (or $p^+$-$n$-$p^+$) configurations and turns the device ON in the $n^+$-$n$-$n^+$ (or $p^+$-$p$-$p^+$) settings. This electron focusing approach to switching based on G*pn*Js holds the exciting possibility of very low energy operation as it can overcome the fundamental thermodynamic limit for subthreshold conduction in electrical current switches: $k_BT\ln10$/decade (where $k_B$ is the Boltzmann constant and $T$ is temperature).

At a G*pn*J, the scattering of the individual pseudospins leads to an electron transmission that is analogous to an electro-magnetic wave incident at a dielectric interface (electron optics) [10]. Electrons with large angles of incidence at the higher electrostatically doped side are unable to conserve their transverse quasi-momentum components across the junction and thus undergo total internal reflection. Those within a critical angle transmit through the tunnel barrier to the opposite side, diverging in the case of unipolar junctions (i.e., $n^+n$ or $p^+p$ junctions) or converging in the case of bipolar junction (*pn*Js) in a manner similar to light passing through convex and concave lenses, respectively. Thus, in the local gating configurations, graphene *pn*J interface behaves as an analogue of the interface of negative index materials. With the proper geometry of G*pn*J's, such electron "optics" allow the conductivity of a graphene sheet to be strongly modulated. Experimental confirmation of the negative index refraction in G*pn*J's has been realized [10], yet demonstration of a high-performance GKTFETs remains elusive [11]. It is postulated that for GKTFETs to have good performance, highly selective angular filtering of carriers must occur at the *pn*Js interface, and specular reflection must

Corresponding author: Son T. Le; Email: son.le@nist.gov

occur at the device's physical edges. In other words, both the electrostatic *pn*Js and the physical edges of the devices need to behave nearly ideally [3].

Understanding the properties of graphene electrostatic *pn*Js has been the subject of numerous studies in recent years with significant progress [12,13,14]. However, even with carefully engineered electrostatic *pn*Js, non-specular scattering at the physical edge of the graphene can compromise the overall performance of GKTFETs [15]. Consequently, understanding the scattering properties of graphene physical edges is crucially important to the engineering of G*pn*J-based switches. Currently, a physical picture of charge scattering at and along the graphene physical edge is not fully developed. To this end, we present here a magnetotransport study of G*pn*Js devices in the quantum Hall regime that elucidates both the scattering properties of G*pn*Js and its physical edges.

In this letter, we present the results of a study of Landau level (LL) edge-state interaction along the physical edge of the graphene in a fully symmetric, ring-shape graphene quantum Hall device with *pn*Js formed by using two independent electrostatic back-gates, measured by magneto-transport. This device geometry allows us to explore the edge-state transport at the physical edge of graphene and provides insight into the nature of edge-state equilibration and carrier transport in graphene.

Previously, by using a specially prepared $SiO_2/Si$ substrate that contains a pair of doped poly-silicon buried gates, or split gates, to form a graded graphene *pn*J [16], we observed that, at the *pn*J interface in the quantum Hall regime, only the edge-state channels of the energetically lowest LLs fully equilibrate across the *pn*J, while the other edge-states remain decoupled and do not transmit current across the junction, the results were later confirmed numerically [17]. The ability to decouple LL edge-states at the *pn*J offers the unique opportunity to probe edge-state interaction at the graphene physical edge where, with an appropriate device

Corresponding author: Son T. Le; Email: son.le@nist.gov

geometry, the edge-state equilibration (or lack thereof) will produce a drastic difference in the overall electric behavior of the device.

The device used in this study was fabricated by using a specially prepared $SiO_2/Si$ substrate that contains a pair of doped poly-silicon gates buried in the $SiO_2$ layer 140 nm from the air-$SiO_2$ interface with an approximately 200 nm gap between the gates. Chemical-mechanical planarization of the substrate achieves an atomically smooth dielectric surface [18]. A dry transfer method [19,20] was used to fabricate and transfer an exfoliated *h*BN/graphene/*h*BN heterostructure onto the buried-gates substrate. The heterostructure was then patterned into a modified, ring-shape Hall bar geometry by using electron-beam lithography and subsequent reactive ion etching. A second e-beam lithography step, followed by the deposition of 5 nm/10 nm/55 nm Cr/Pd/Au and subsequent metal lift-off, was used to form 1D metal edge contacts to the *h*BN-sandwiched graphene [19]. The ring-shape geometry, combined with two local gates, creates two *pn*Js with identical electrostatic profiles. The regions of the device above local gates G1 and G2 are referred to as region 1 and region 2, respectively. Fig. 1a shows a schematic 3D rendering of the complete device with Fig. 1b showing the SEM cross-section of the buried-gate structure with a superimposed representation of the cross-section of the edge-contacted *h*BN/graphene/*h*BN device at the middle. Figure 1c's inset is an atomic force microscopy (AFM) image of the final device.

Measurements were performed at a base temperature of approximately 2 K and a magnetic field of 12 T by using a conventional four-terminal transport measurement with lock-in amplifiers set to an oscillator frequency of 47 Hz with a current bias of approximately 500 nA. Voltages applied to the two poly-silicon split gates, $V_{G1}$ and $V_{G2}$, allow independent tuning of both carrier type and concentration in the two halves of the graphene ring. The measurement range of $V_{G1}$ and $V_{G2}$ was set such that ($|V_{G1}|$, $|V_{G2}|$, $|V_{G2}$-$V_{G1}|$) ≤ 28 V) to minimize gate leakage current while,

Corresponding author: Son T. Le; Email: son.le@nist.gov

at the same time, maximizing the observable filling factor range. The charge neutrality point, *CNP* = -2V, is close to zero gate voltage, indicative of high quality, low intrinsically doped graphene. The mobility of the device at $T$ = 2 K with a carrier density of $n \approx 5\text{x}10^{11}$ cm$^{-2}$ was measured to be approximately 40,000 cm$^2$/Vs, estimated from the method described in Ref. 21.

To study LL edge-state interactions in the quantum Hall regime, we measured the ring's resistance as a function of electrostatic doping in the two halves of the ring. When a strong perpendicular magnetic field is applied to the device, holes and electrons in, respectively, *p*- and *n*-regions of the device transition into quantum Hall states. Bulk electrons and holes are strongly localized by the magnetic field, causing the device transport properties to be governed by the formation of chiral LL edge-state channels circulating along the boundary and the *pn*J interface. The edge channels circulate in opposite directions for *p*- and *n*-regions, which results in edge states of opposite charge traveling in the same direction, alongside each other, at the electrostatic *pn*J interface (Fig. 2c). In the quantum Hall regime, $\nu_1$, $\nu_2$ are the filling factors in each region (for graphene $\nu_1$, $\nu_2$ = ±2, ±6, ±10…). To probe the LL edge-state interactions at the graphene physical edge in our device, non-equilibrium conditions of the local chemical potential of the LL edge-states were established by decoupling the edge-state of the energetically lowest LL from the energetically higher LLs at one *pn*J [16]. Due to the chirality of these edge channels, which is determined by the direction of the magnetic field and the charge of the carriers (*n*- or *p*-type), they then pass adjacent to the abrupt physical inner edge of the graphene ring before reaching the second *pn*J which "senses" the amount of equilibration that occurred between the lowest LL and the others. The smooth curvature of the ring avoids possible geometric mixing at sharp physical corners.

Corresponding author: Son T. Le; Email: son.le@nist.gov

A 2D color map projection of the measured longitudinal resistance $R_{14,23}$ [22] across the ring as a function of $V_{G1}$ and $V_{G2}$, both swept independently from -16 V to +12 V (corresponding to filling factors up to ±10) is shown in Figure 2a for $B$ = 12 T. The resistance map, where the numbers in each block correspond to the approximate resistance plateau values in units of $h/e^2$, reveals a complex picture of quantized resistance plateaus that includes fractional values of $h/e^2$. For example, the lower-right quadrant of the 2D map (*np*J: $V_{G1} > 0$ and $V_{G2} < 0$) contains 1 $h/e^2$, 3/4 $h/e^2$, and ≈ 3/5 $h/e^2$ plateaus; other areas of the resistance map show plateaus of 1/15 $h/e^2$, 1/3 $h/e^2$, 2/5 $h/e^2$ and ≈ 4/15 $h/e^2$ (curve (*ii*) and (*iii*) in Fig. 1c) which are located symmetrically about the $V_{G1} = -V_{G2}$ line.

The conditions where the two halves of the graphene ring are at the same polarity (*pp*+ and *nn*+) correspond to the upper-right (*nn*J: $V_{G1} > 0$ and $V_{G2} > 0$) and lower-left (*pp*J: $V_{G1} < 0$ and $V_{G2} < 0$) quadrants of the $R_{14,23}$ resistance map in Fig. 2a. For these cases, the mechanism of quantized resistance plateau formation can be explained by using the Landauer-Büttiker edge-state formalism [23] with no further assumptions regarding equilibration or non-equilibration of the edge-states [24]. The edge-states that can be accommodated in the lower carrier density region circulate around the entirety of the device, including both regions 1 and 2. Additional edge-states arise in the higher carrier concentration region that do not cross the *pn*J interface and circulate only in the region of higher filling factors [25]. Under these conditions, the same equations are obtained for the longitudinal quantized resistance plateaus independent of the equilibration or lack thereof between the edge-states, and are calculated as follows:

For a unipolar *nn*J, *n*-type in both regions 1 and 2:

$$R_{14,23}\left(h/e^2\right)=\begin{cases}0, & |\nu_1|>|\nu_2|\\ |\nu_1-\nu_2|/|\nu_1\nu_2|, & |\nu_1|<|\nu_2|\end{cases} \tag{1a}$$

Corresponding author: Son T. Le; Email: son.le@nist.gov

For a unipolar *pp*J, *p*-type in both regions 1 and 2:

$$R_{14,23}\left(h/e^2\right)=\begin{cases}\left|\nu_1-\nu_2\right|/\left|\nu_1\nu_2\right|, & \left|\nu_1\right|>\left|\nu_2\right| \\ 0, & \left|\nu_1\right|<\left|\nu_2\right|\end{cases} \tag{1b}$$

These calculated $R_{14,23}$ values are shown for the *nn*J-regime and the *pp*J-regime in, respectively, the upper-right and the lower-left quadrants of the 2D resistance maps of $R_{14,23}$ evaluated for two different possible cases of edge-state equilibration in Figs. 2b and 2d (to be discussed later in this paper). The calculated $R_{14,23}$ values match well with the measured 2D resistance maps of $R_{14,23}$ shown in Fig. 2a.

In the case of a bipolar (*p*-*n*) regime, represented by the upper-left and lower-right quadrants of the map in Fig. 2a, the edge-states in the *p*- and *n*-regions circulate in opposite directions along the outer and inner edges of the graphene ring and thus adjacent along the *pn*Js as illustrated in Fig. 2c. We have previously confirmed that in such graphene devices only the lowest LL's edge-state equilibrates at the *pn*J [16], this local imbalance gives us a unique opportunity to probe edge-state interaction along the inner edge of the ring where the interaction of the higher LLs with the lowest LL along the physical edge will determine the overall quantized conductance behavior of the device.

Consider the case of the device in a strong positive magnetic field in the bipolar regime, with the region gated by G1 tuned to p-type polarity and the region gated by G2 tuned to n-type polarity (see figure 2c). In the quantum Hall regime, hole LL edge-states in the p-type half of the device will travel in the clockwise direction while electron LL edge-states in the n-type half of the device will travel in the counter-clockwise direction. The electron and hole LL edge-states will first meet each other at *pn*J1, called the *conditioning junction*. At the *pn*J1 interface, non-equilibrium conditions of the local chemical potential of the LL edge-states are established by decoupling the edge-state of the energetically lowest LL from the

Corresponding author: Son T. Le; Email: son.le@nist.gov

higher LLs in both electron (G1) and hole (G2) side of the device. The edge-states, with the lowest LL edge-state initially carrying a different chemical potential from the higher LL edge-states, will then travel along the inner physical edge of the device: hole LL edge-states will travel along the p-type half of the inner physical edge and electron LL edge-states will travel along the n-type half of the inner physical edge. Along this inner physical edge of the device, the interaction of the lowest LL edge-state and the higher LL edge-states will alter the local chemical potential of the LL edge-states in the following possible ways, which we will refer to as case 1 and case 2, respectively. If, in case 1, the lowest LL edge-state does not equilibrate with the higher LL edge-states along the physical inner edge of the device, those LL edge-states will arrive at *pn*J2, the *sensing junction*, carrying the same chemical potential they acquired at the conditioning junction, *pn*J1. Alternatively, in case 2, if the lowest LL edge-state equilibrates with the higher LL edge-states along the physical inner edge, those edge-states will come to *pn*J2 carrying the same chemical potential, the value of which is the weighted average of the chemical potential of the lowest LL edge-state and the higher LL edge-states. Finally, the sensing junction, *pn*J2, will, again, decouple the lowest LL edge-state from the higher LL edge-states by equilibrating only the lowest LL edge-state across the *pn*J, after which point the charge carriers will travel to the electrical contacts and the LL edge-state interactions in the device can be assessed by means of a measurement of the device resistance. It is important to note that this "sensing mechanism" only works if both the *pn*J1 and *pn*J2 decouple the lowest LL edge-states from the higher LL edge-states. If any or both of the two *pn*Js equilibrate the LL edge-states at their respective *pn*J interfaces, then the device will behave the same as the device presented in [24] regardless of the interaction of LL edge-states at the physical inner edge of the device.

Corresponding author: Son T. Le; Email: son.le@nist.gov

Based on our previously developed model, and using Landauer-Büttiker edge-state formalism, we calculate the expected resistance values of our ring-shape device as a function of filling factor in two cases: (1) Assuming no equilibration of LL edge-states along the inner edge of our graphene ring (Eq. 2); and (2) Assuming full equilibration of LL edge-states along the inner edge of the ring (Eq. 3).

$$R_{14,23}\left({}^{h}\!/_{e^2}\right) = \begin{cases} \frac{(|\nu_1|-2)}{2|\nu_1|} + \frac{(|\nu_2|-2)}{2|\nu_1|} & p\text{- type region } 1, n\text{- type region } 2 \\ 1 & n\text{- type region } 1, p\text{- type region } 2 \end{cases} \quad (2)$$

$$R_{14,23}\left({}^{h}\!/_{e^2}\right) = \begin{cases} \frac{(|\nu_1\nu_2|-|\nu_1|-|\nu_2|)^2}{|\nu_1\nu_2|(2|\nu_1\nu_2|-|\nu_1|-|\nu_2|)} & p\text{- type region } 1, n\text{- type region } 2 \\ \frac{|\nu_1\nu_2|}{2|\nu_1\nu_2|-|\nu_1|-|\nu_2|} & n\text{- type region } 1, p\text{- type region } 2 \end{cases} \quad (3)$$

Fig. 2b shows the calculated 2D resistance map of $R_{14,23}$ for the case 1 assumption for the *pn*J and *np*J regimes (upper-left and lower-right quadrants, respectively) per Eq. 2. Fig. 2c illustrates the edge-state model based on the case 2 assumption, and Fig. 2d shows the calculated 2D resistance map of $R_{14,23}$ for this case for the *pn*J and *np*J regimes (upper-left and lower-right quadrants, respectively) per Eq. 3. In the case of Fig. 2b, calculated for the case of no equilibration of edge-states along the inner edge of our graphene ring, the ring-shaped device is shown to behave like a linear *pn*J device [16]. It is clear that the assumption of no equilibration along the physical edge of the graphene fails to explain all measured $R_{14,23}$ resistance plateaus in bipolar regime in Fig. 2a. On the other hand, the full equilibration assumption explains the measured resistance values very well, signifying complete equilibration between edge channels over a rather short approximately 1 µm distance (≈).

In our previous work, a low-mobility device was studied in an effort induce the greatest possible scattering and equilibrium at the *pn*J interface. Nevertheless,

Corresponding author: Son T. Le; Email: son.le@nist.gov

we observed only the lowest LL's edge-state equilibrated at the *pn*J interface due to the smooth electrostatic potential profile of the *pn*J. In this study, we have used a relatively high-mobility hBN/graphene/hBN sample in an effort to minimize scattering and equilibrium at the electrostatics *pn*Js and ensure that the two *pn*J "sensing" scheme works effectively. The observation of full equilibration of LLs edge-states at the sample's edges reveals that the interaction of LL edge-states along the physical edge is very strong in contrast to the lack of equilibration along an electrostatic *pn*J. Qualitatively, it should be much easier for the LLs edge-states to equilibrate along the graphene physical edge in comparison with the electrostatically defined interface due to the abruptness and roughness nature of the physical edge. The physical edge of graphene can be considered as an infinitely high potential barrier with an atomically abrupt potential profile at the graphene/vacuum boundary. Due to this sharp barrier the LL edge-states are physically close to each other allowing strong interactions. In addition, because the physical edges are formed by reactive ion etching of the initial graphene flake into the lithographically defined shape, there are significant electronic defects and physical roughness at the graphene/vacuum interface which leads to a high scattering rate between the closely spaced LL edge-states. An abrupt potential profile and high carrier scattering rates have been theoretically predicted to promote strong LL edge-state mixing at *pn*J interfaces [17,26] and the same mixing processes are expected to hold true at the device's physical edge.

In conclusion, we have characterized the interaction of LL edge-states at the graphene physical edge by using a ring-shaped graphene device with a pair of *pn*Js. Our measurements confirm that the energetically lowest LL edge-state is decoupled from the others, and that only this lowest LL edge-state couples across the electrostatic junction and contributes to total current. In contrast, the LL edge-states strongly interact and fully equilibrate at the graphene physical edge despite the

Corresponding author: Son T. Le; Email: son.le@nist.gov

relatively high carrier mobilities and short graphene edge distance (~1 um) in the studied samples. The strong LL edge-state interactions arise from the abrupt and rough nature of the device's physical edges. Understanding the underlying physics of the interaction between LL edge-states at graphene physical edges leads to a better understanding of carrier scattering at a graphene device's boundary which is fundamental to overcome the challenges of engineering future high-performance graphene field-effect transistors based upon electron optics. Furthermore, the quantum Hall edge states in graphene, if coupled to an s-wave superconductor by proximity effect, are expected to form non-Abelian chiral anyon modes [27,28] that can allow a graphene quantum Hall-superconductor system to be used in the realization of a topological quantum computer. The manipulation of topologically protected chiral edge modes, such as by utilizing the coupling and decoupling of LL edge-states at the physical and electrostatic edges of graphene discussed here, is essential to the quantum computer that utilizes braiding of topological chiral edge states for computation [29]. The finding also offers us a new degree of freedom to controllably tune quantized resistance values for the application of graphene-based scalable resistance standards by combining *pn*Js with properly designed device geometries.

**Acknowledgements**

S.T.L. acknowledges support by the National Institute of Standards and Technology (NIST) grant 70NANB16H170. Research performed in part at the NIST Center for Nanoscale Science and Technology nanofabrication facility.

Corresponding author: Son T. Le; Email: son.le@nist.gov

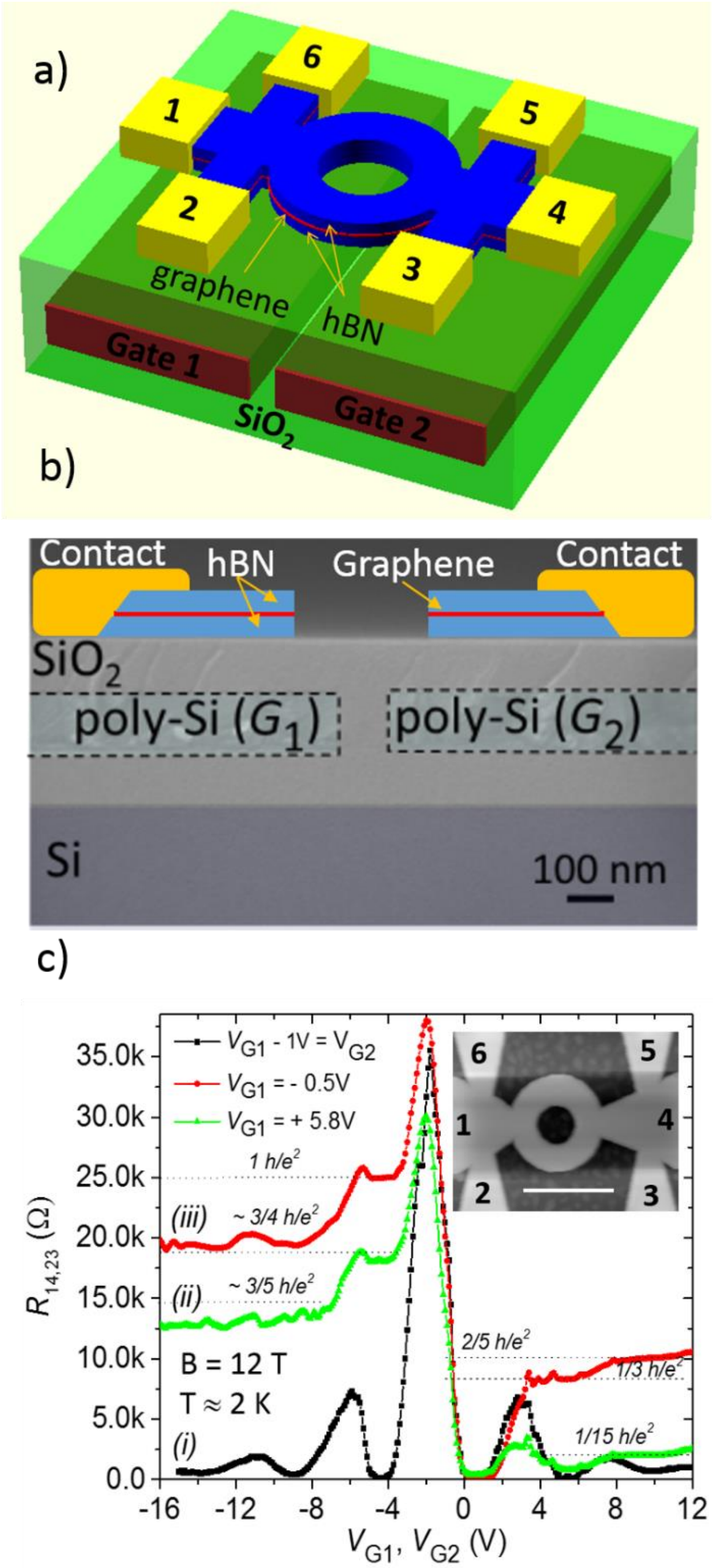

*Figure 1: (a) 3-dimenstional rendering of the device schematic showing complete device structure with split, buried gates. (b) SEM image of the substrate's cross-section with buried poly-silicon gates in $SiO_2$ (scale bar is 100 nm) and a superimposed schematic of the cross-section of the device at the middle. (c) Longitudinal resistance $R_{14,23}$ measured at T = ~2 K and B = 12 T as a function of local gate voltages $V_{G1}$ and $V_{G2}$; (i) $V_{G1}$ -1V= $V_{G2}$, (ii) $V_{G1}$ = +5.8 V, and (iii) $V_{G1}$ = -0.5 V. (c)'s inset: AFM image of ring-shape graphene pnJ device in a Hall bar geometry, scale bar is 1 µm.*

Corresponding author: Son T. Le; Email: son.le@nist.gov

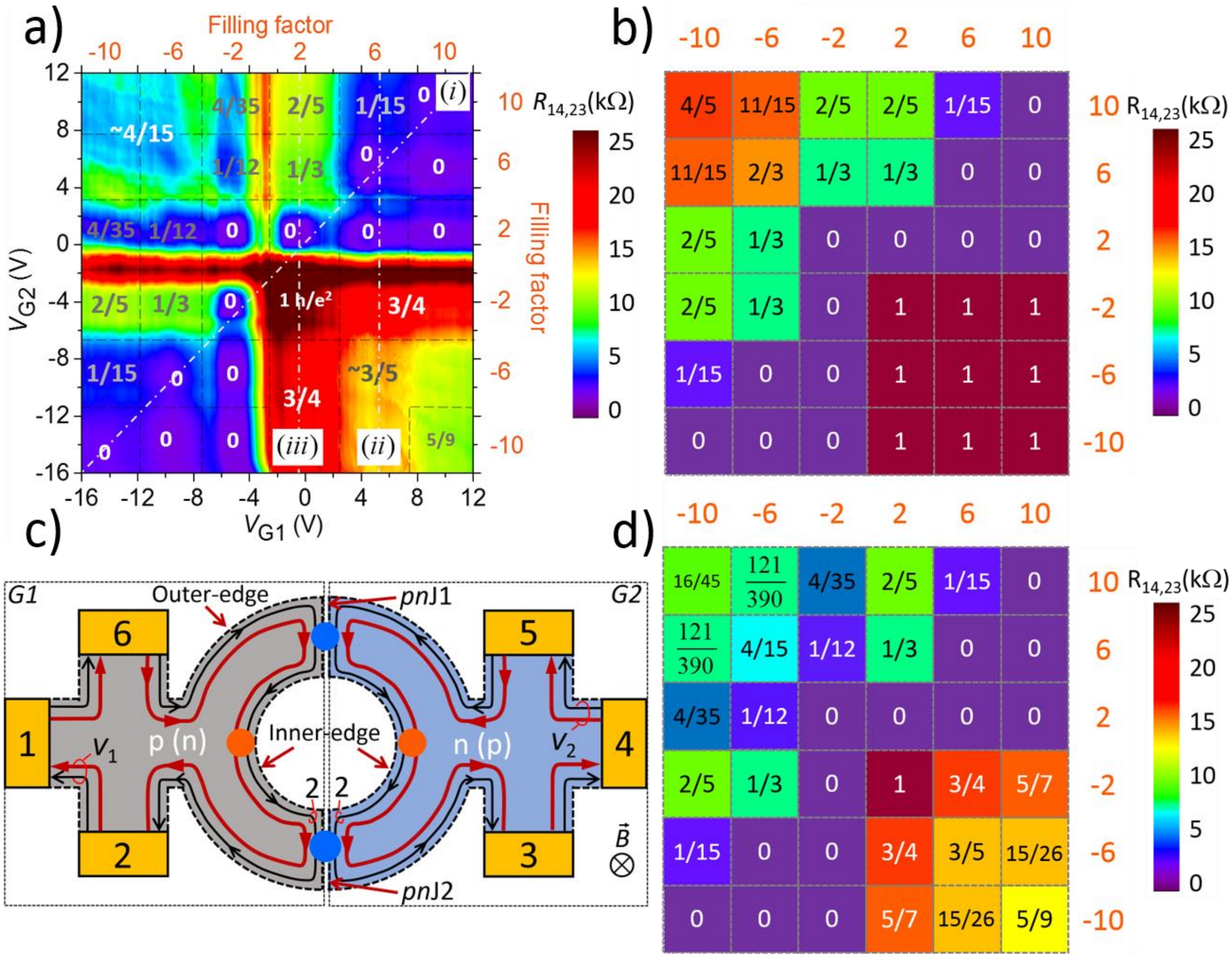

*Figure 2: (a) Color maps 2D projection of longitudinal resistance $R_{14,23}$ measured at $T$ = 2 K and $B$ = 12 T as a function of local gate voltages ($V_{G1}$, $V_{G2}$) and filling factors ($\nu_1$, $\nu_2$). Numbers inside the map are plateau resistance values in units of $h/e^2$. The dash-dotted lines belong to (i)-, (ii)-, and (iii)-curves in Figure 1(c). (b) Calculated 2D resistance map of $R_{14,23}$ as a function of filling factors ($\nu_1$, $\nu_2$) with lowest Landau level edge-state equilibration at the pnJs and assuming no equilibration of edge-states at the graphene ring inner edges. (c) Schematic of edge-state transport model of our ring-shaped hBN/G/hBN device in the quantum Hall regime, the edge-states in regions 1 and 2 are associated with filling factors $\nu_1$ and $\nu_2$, respectively. In this model, only the lowest Landau level edge-states on opposite sides of the pnJs (marked by thin black lines) equilibrate at the pnJs interface (blue dots) while the energetically higher LL edge-states (marked by thick red lines) are not equilibrated at the pnJs interface but fully equilibrated with the lowest Landau level at the inner edge of the graphene ring (orange dots). Blue dots and orange dots indicate the region of edge-state equilibration (dot's positions are schematic and not a detailed representation of carrier dynamics at the graphene pnJ interface and edges). (d) Calculated 2D resistance map of $R_{14,23}$ as a function of filling factors ($\nu_1$, $\nu_2$) based on the model presented in (c), numbers inside the map are plateau resistance values in units of $h/e^2$.*

Corresponding author: Son T. Le; Email: son.le@nist.gov

Corresponding author: Son T. Le; Email: son.le@nist.gov

Corresponding author: Son T. Le; Email: son.le@nist.gov

Corresponding author: Son T. Le; Email: son.le@nist.gov